\documentclass[showpacs,preprintnumbers,amsmath,amssymb,prl,aps,superscriptaddress,doublespace,twocolumn]{revtex4-1}
\usepackage{graphicx}
\usepackage{amsmath}
\usepackage{verbatim}
\setcitestyle{super}

\newcommand{\kpsii}{\ket{\psi^0}}
\newcommand{\bpsii}{\bra{\psi^0}}
\newcommand{\Fp}{F_{\mathrm{p}}}

\newcommand{\OddBellp}{ \ket{\Phi^{+}}}
\newcommand{\EvenBellp}{ \ket{\Psi^{+}}}

\newcommand{\BraOddBellp}{ \bra{\Phi^{+}}}

\newcommand{\epsp}{\epsilon_\mathrm{p}}
\newcommand{\epse}{\epsilon_\mathrm{e}}
\newcommand{\epso}{\epsilon_\mathrm{o}}
\newcommand{\tp}{\tau_\mathrm{P}}

\newcommand{\Vp}{V_\mathrm{P}}
\newcommand{\rhoq}{\rho}
\newcommand{\rhoi}{\rho^{(0)}}

\newcommand{\rmixed}{\rho_{11,10}}
\newcommand{\rodd}{\rho_{01,10}}
\newcommand{\reven}{\rho_{00,11}}

\newcommand{\kebits}{{\rm kebits/s}}

\newcommand{\Ratee}{R_{\mathrm{e}}}
\newcommand{\Rateexpt}{R_{\mathrm{exp}}}

\newcommand{\ket}[1]{\left\lvert #1 \right\rangle}
\newcommand{\bra}[1]{\left\langle #1 \right\rvert}
\newcommand{\avg}[1]{\left\langle #1 \right\rangle}
\newcommand{\QA}{\mathrm{Q}_{\mathrm{A}}}
\newcommand{\QB}{\mathrm{Q}_{\mathrm{B}}}
\newcommand{\K}{\mathrm{K}}

\newcommand{\fa}{f_\mathrm{A}}
\newcommand{\fb}{f_\mathrm{B}}

\newcommand{\fp}{f_\mathrm{p}}

\newcommand{\Vint}{V_\mathrm{int}}

\newcommand{\ti}{t_\mathrm{i}}
\newcommand{\tf}{t_\mathrm{f}}
\newcommand{\us}{\mu\mathrm{s}}
\newcommand{\ns}{\mathrm{ns}}
\newcommand{\tr}{\text{Tr}}
\newcommand{\Ojoint}{\mathcal{O}}
\newcommand{\betaA}{\beta_{\mathrm{A}}}
\newcommand{\betaB}{\beta_{\mathrm{B}}}
\newcommand{\betaBA}{\beta_{\mathrm{BA}}}
\newcommand{\ToneA}{T_{1}^\mathrm{A}}
\newcommand{\ToneB}{T_{1}^\mathrm{B}}
\newcommand{\TtwoA}{T^{\varphi,\mathrm{A}}_{2}}
\newcommand{\TtwoB}{T^{\varphi,\mathrm{B}}_{2}}
\newcommand{\Ttwoq}{T^{\varphi,q}_2}
\newcommand{\smA}{\sigma_-^\mathrm{A}}
\newcommand{\smB}{\sigma_-^\mathrm{B}}
\newcommand{\spA}{\sigma_+^\mathrm{A}}
\newcommand{\spB}{\sigma_+^\mathrm{B}}
\newcommand{\szA}{\sigma_z^\mathrm{A}}
\newcommand{\szB}{\sigma_z^\mathrm{B}}
\newcommand{\szq}{\sigma_z^q}

\newcommand{\ChiA}{\chi_{\mathrm{A}}}
\newcommand{\ChiB}{\chi_{\mathrm{B}}}
\newcommand{\kHz}{\mathrm{kHz}}
\newcommand{\MHz}{\mathrm{MHz}}
\newcommand{\GHz}{\mathrm{GHz}}
\newcommand{\nss}{\bar{n}_{\mathrm{ss}}}
\newcommand{\dB}{\mathrm{dB}}
\newcommand{\Mp}{M_\mathrm{P}}
\newcommand{\Vthreshp}{V_{\mathrm{th}+}}
\newcommand{\Vthreshm}{V_{\mathrm{th}-}}
\newcommand{\Vthresh}{V_{\mathrm{th}}}
\newcommand{\conc}{\mathcal{C}}
\newcommand{\LN}{E_\mathcal{N}}
\newcommand{\psuccess}{p_{\mathrm{success}}}
\newcommand{\rateE}{\mathcal{R}_{e}}
\newcommand{\omegar}{\omega_{\mathrm{r}}}
\newcommand{\omegap}{\omega_{\mathrm{p}}}
\newcommand{\omegaq}{\omega_q}
\newcommand{\chiQ}{\chi_q}
\newcommand{\dt}{\mathrm{d}t}
\newcommand{\out}[2]{{\ket{#1}\bra{#2}}}

\newcommand{\expec}[1]{\langle \: #1 \:\rangle}
\newcommand{\com}[2]{ \left[ #1,#2 \right]}
\newcommand{\acom}[2]{ \{ #1,#2 \}}

\begin{document}
\title{Deterministic entanglement of superconducting qubits by\\ parity measurement and feedback}

\author{D.~Rist\`e}

\author{M.~Dukalski}

\author{C.~A.~Watson}

\author{G.~de Lange}

\author{M.~J.~Tiggelman}

\author{Ya.~M.~Blanter}
\affiliation{Kavli Institute of Nanoscience, Delft University of Technology, P.O. Box 5046,
2600 GA Delft, The Netherlands}

\author {K.~W.~Lehnert}
\affiliation{JILA, National Institute of Standards and Technology and Department of Physics, University of Colorado, Boulder, Colorado 80309, USA}
 
\author{R.~N.~Schouten}

\author{L.~DiCarlo}
\affiliation{Kavli Institute of Nanoscience, Delft University of Technology, P.O. Box 5046,
2600 GA Delft, The Netherlands}

\date{\today}

\begin{abstract}
The stochastic evolution of quantum systems during measurement is arguably the most enigmatic feature of quantum mechanics. 
Measuring a quantum system typically steers it towards a classical state, destroying any initial quantum superposition and any entanglement with other quantum systems. 
Remarkably, the measurement of a shared property between non-interacting quantum systems can generate entanglement starting from an uncorrelated state. Of special interest in quantum computing is the parity measurement~\cite{Ruskov03}, 
which projects a register of quantum bits (qubits) to a state with an even or odd total number of excitations. Crucially, a parity meter must discern the two parities with high fidelity while preserving coherence between same-parity states.  
Despite numerous proposals for atomic~\cite{Kerckhoff09}, semiconducting~\cite{Ruskov03,Engel05, Trauzettel06, Ionicioiu07, Williams08b, Haack10}, and superconducting qubits~\cite{Lalumiere10,Tornberg10}, realizing a parity meter creating entanglement for both even and odd measurement results has remained an outstanding challenge. 
We realize a time-resolved, continuous parity measurement of two superconducting qubits using the cavity in a 3D circuit quantum electrodynamics (cQED)~\cite{Blais04,Paik11} architecture and phase-sensitive parametric amplification~\cite{Castellanos-Beltran08}. 
Using postselection, we produce entanglement by parity measurement reaching $77\%$ concurrence. Incorporating the parity meter in a feedback-control loop, we transform the entanglement generation from probabilistic to fully deterministic, achieving $66\%$ fidelity to a target Bell state on demand. These realizations of a parity meter and a feedback-enabled deterministic measurement protocol provide key ingredients for active quantum error correction in the solid state~\cite{Nielsen00,Ahn02,Devoret13}. 
\end{abstract}

\maketitle

Recent advances in nearly quantum-limited amplification~\cite{Castellanos-Beltran08} and improved qubit coherence times in 3D cQED~\cite{Paik11} have allowed the first investigations of the gradual collapse of single-qubit wavefunctions in the solid state~\cite{Hatridge13,Murch13}, on par with previous fundamental studies in atomic  systems~\cite{Guerlin07}. 
The continuous measurement of a joint property extends this study to the multipartite setting, resolving the projection to states which are inaccessible via individual qubit measurements.  
In a two-qubit system, the ideal  parity measurement transforms an unentangled superposition state $\kpsii = (\ket{00}+\ket{01}+\ket{10}+\ket{11})/2$ into  Bell states
\begin{eqnarray*}
&&\OddBellp = \frac{1}{\sqrt{2}}(\ket{01}+\ket{10})\,\,\, \mathrm{and} \,\,\, \EvenBellp = \frac{1}{\sqrt{2}}(\ket{00}+\ket{11})
\end{eqnarray*}
for odd and even outcome, respectively. 
Beyond generating entanglement between non-interacting qubits~\cite{Ruskov03, Trauzettel06, Ionicioiu07, Williams08b, Haack10}, parity measurements allow deterministic two-qubit gates~\cite{Beenakker04, Engel05} and play a key role as syndrome detectors in quantum error correction~\cite{Nielsen00,Ahn02}. 
A heralded parity measurement has been recently realized for nuclear spins in diamond~\cite{Pfaff12}. 
By minimizing measurement-induced decoherence at the expense of single-shot fidelity, highly entangled states were generated with $3\%$ success probability. 
Here, we realize the first solid-state parity meter that produces entanglement with unity probability. 

\begin{figure}
\includegraphics[width=0.94\columnwidth]{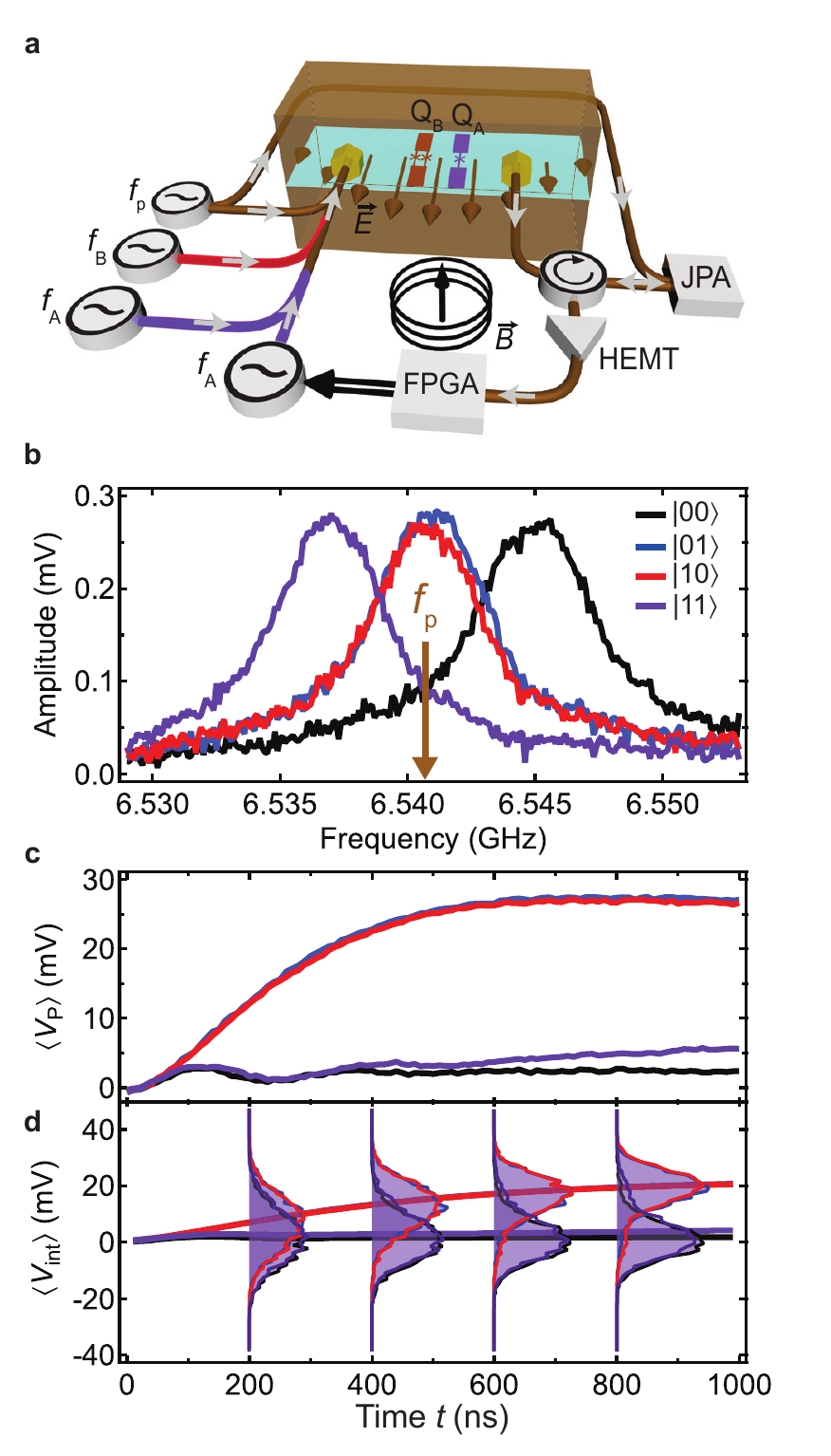} 
\caption{{\bf Realization of cavity-based two-qubit parity readout in circuit QED.} {\bf a,} Simplified diagram of the experimental setup. Single- and double-junction transmon qubits ($\QA$ and $\QB$, respectively) dispersively couple to the fundamental mode of a 3D copper cavity enclosing them. Parity measurement is performed by homodyne detection of the qubit state-dependent cavity response~\cite{Blais04} using phase-sensitive Josephson parametric amplification (JPA)~\cite{Castellanos-Beltran08}. Following further amplification at $4~\K$ (HEMT) and room temperature, the signal is demodulated and integrated. A field-programmable-gate-array (FPGA) controller closes the feedback  loop that achieves deterministic entanglement by parity measurement (Fig.~4). See  Fig.~S2 for a detailed schematic of the setup. {\bf b,} Matching of the dispersive cavity shifts realizing a parity measurement.  
 {\bf c,} Ensemble-averaged homodyne response $\avg{\Vp}$ for qubits prepared in the four computational basis states. 
 {\bf d,} Curves: corresponding ensemble averages of the running integral $\avg{\Vint}$ of $\avg{\Vp}$ between $\ti=0$ and $\tf=t$. Single-shot histograms ($5,000$ counts each) of $\Vint$ are shown in $200~\ns$ increments. %, with $1.2~\mV$ bin size. 
} 
\end{figure}

Our parity meter realization exploits the dispersive regime~\cite{Blais04} in two-qubit cQED. Qubit-state dependent shifts of a cavity resonance (here, the fundamental of a 3D copper cavity enclosing transmon qubits $\QA$ and $\QB$) allow joint qubit readout by homodyne detection of an applied microwave pulse transmitted through the cavity (Fig.~1a).   
The temporal average $\Vint$ of the homodyne response $\Vp(t)$ over the time interval $[\ti,\tf]$ 
constitutes the measurement needle, with expectation value
\[
\langle \Vint \rangle = \tr(\Ojoint \rho),
\]
where $\rho$ is the two-qubit density matrix and the observable $\Ojoint$ has the general form
\[
\Ojoint = \beta_0  + \betaA   \szA + \betaB \szB  + \betaBA \szB  \szA.
\]
The coefficients $\beta_0$, $\betaA$, $\betaB$, and $\betaBA$  depend on the strength $\epsp$, frequency $\fp$ and duration $\tp$ of the measurement pulse, the cavity linewidth $\kappa$, and the frequency shifts $2\ChiA$ and $2\ChiB$ of the fundamental mode when $\QA$ and $\QB$ are individually excited from $\ket{0}$ to $\ket{1}$. 
The necessary condition for realizing a parity meter is $\betaA=\betaB=0$ ($\beta_0$ constitutes a trivial offset). A simple approach~\cite{Hutchison09,Lalumiere10}, pursued here, is to set $\fp$ to the average of the resonance frequencies for the four computational basis states  $\ket{ij}$ ($i,j\in\{0,1\}$) and to match $\ChiA=\ChiB$. 
We engineer this matching by targeting specific qubit transition frequencies $\fa$ and $\fb$ below and above the fundamental mode during fabrication and using an external magnetic field to fine-tune $\fb$ in situ (Fig.~S1). We align $\ChiA$ to $\ChiB$ to within $\sim0.06\,\kappa=2\pi \times 90~\kHz$ (Fig.~1b). 
The ensemble-average $\avg{\Vp}$ confirms nearly identical high response for odd-parity computational states $\ket{01}$ and $\ket{10}$, and nearly identical low response for the even-parity $\ket{00}$ and $\ket{11}$ (Fig.~1c). The transients observed are consistent with the independently measured $\kappa$, $\ChiA$ and $\ChiB$ values, and the $4~\MHz$ bandwidth of the Josephson parametric amplifier (JPA) at the front end of the output amplification chain. Single-shot histograms (Fig.~1d) demonstrate the increasing ability of $\Vint$ to discern states of different parity as $\tf$ grows (keeping $\ti=0$), and its inability to discriminate between states of the same parity.  The histogram separations at $\tf=400~\ns$ give $|\betaA|,|\betaB| < 0.02~|\betaBA|$ (Fig.~S3).

\begin{figure}
\includegraphics[width=0.9\columnwidth]{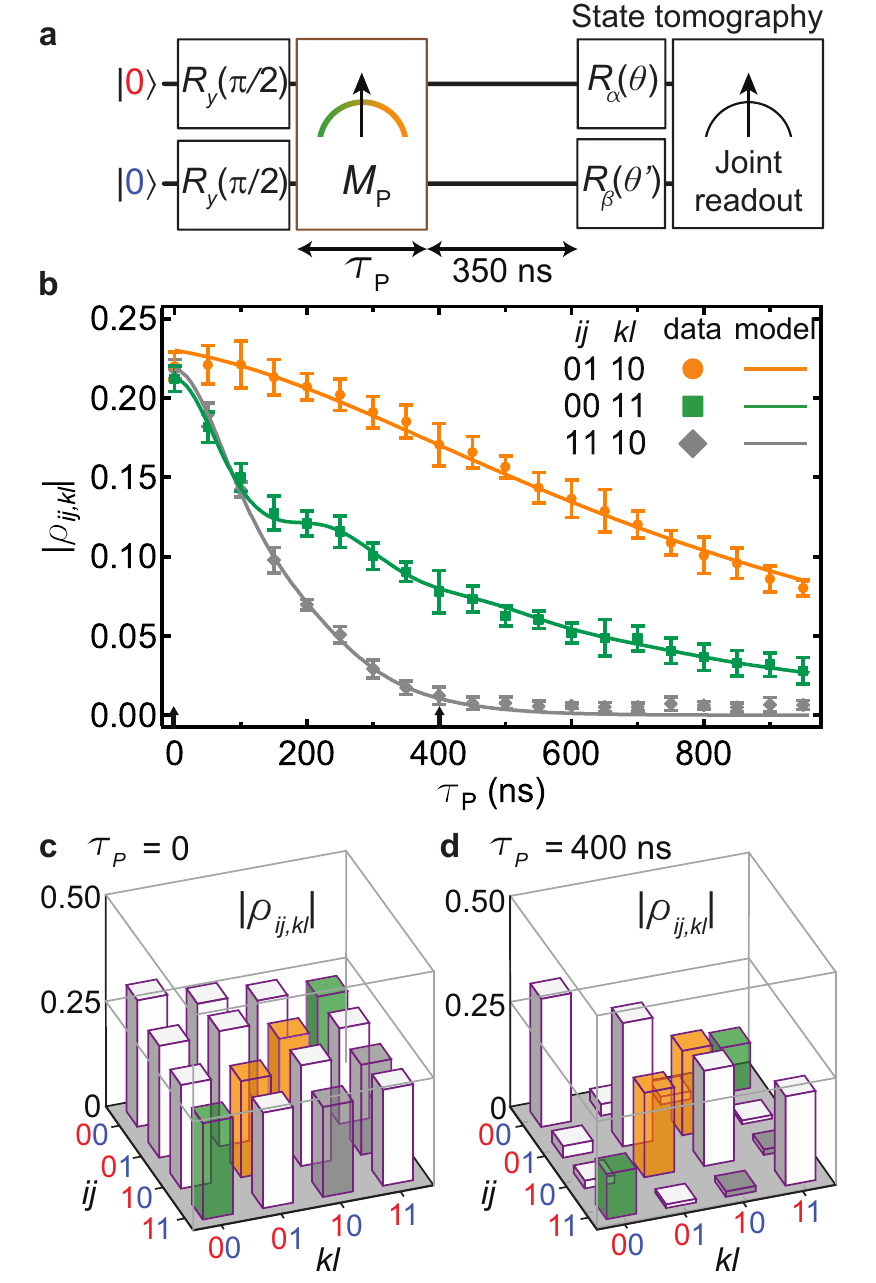}
\caption{{\bf Unconditioned two-qubit evolution under continuous parity measurement.} {\bf a,} Pulse sequence including preparation of the qubits in the maximal superposition state $\rhoi=\kpsii\bpsii$, parity measurement and tomography of the final two-qubit state $\rho$ using joint readout. {\bf b,} Absolute coherences $|\rmixed|$, $|\rodd|$, $|\reven|$ following a parity measurement with variable duration $\tp$. Free parameters of the model are the steady-state photon number on resonance $\nss = 2.5\pm 0.1$, the difference $(\ChiA-\ChiB)/\pi=235\pm4~\kHz$, and the absolute coherence values at $\tp=0$ to account for few-percent pulse errors in state preparation and tomography pre-rotations. Note that the frequency mismatch differs from that in Fig.~1b due to its sensitivity to measurement power (see also Fig.~S6). {\bf c,d,} Manhattan-style plots of extracted density matrices for $\tp=0$ (c) and $\tp=400~\ns$ (d), by which time coherence across the parity subspaces (grey) is almost fully suppressed, while coherence persists within the odd-parity (orange) and even-parity (green) subspaces. Error bars correspond to the standard deviation of $15$ repetitions. See  Fig.~S4 for the temporal evolution with parity measurement off and Figs.~S5, S6 for two-qubit tomography at other values of $\tp$ and $\nss$, respectively.}
\label{Fig_2}
\end{figure}

Moving beyond the description of the measurement needle, we now investigate the collapse of the two-qubit state during parity measurement. We prepare the qubits in the maximal superposition state $\kpsii=\frac{1}{2}\left(\ket{00}+\ket{01}+\ket{10}+\ket{11}\right)$, apply a parity measurement pulse for $\tp$, and perform tomography of the final two-qubit density matrix $\rho$ with and without conditioning on $\Vint$ (Fig.~2a). We choose a weak parity measurement pulse exciting $\nss=2.5$ intra-cavity photons on average in the steady-state, at resonance. A delay of $3.5/\kappa=350~\ns$ is inserted to deplete the cavity of photons before performing  tomography. The tomographic joint readout is also carried out at $\fp$, but with $14~\dB$ higher power, at which the cavity response is weakly nonlinear and sensitive to both single-qubit terms and two-qubit correlations ($\betaA\sim\betaB\sim\betaBA$, see  Fig.~S3), as required for tomographic reconstruction~\cite{Filipp09}.

\begin{figure*}
\includegraphics[width=2\columnwidth]{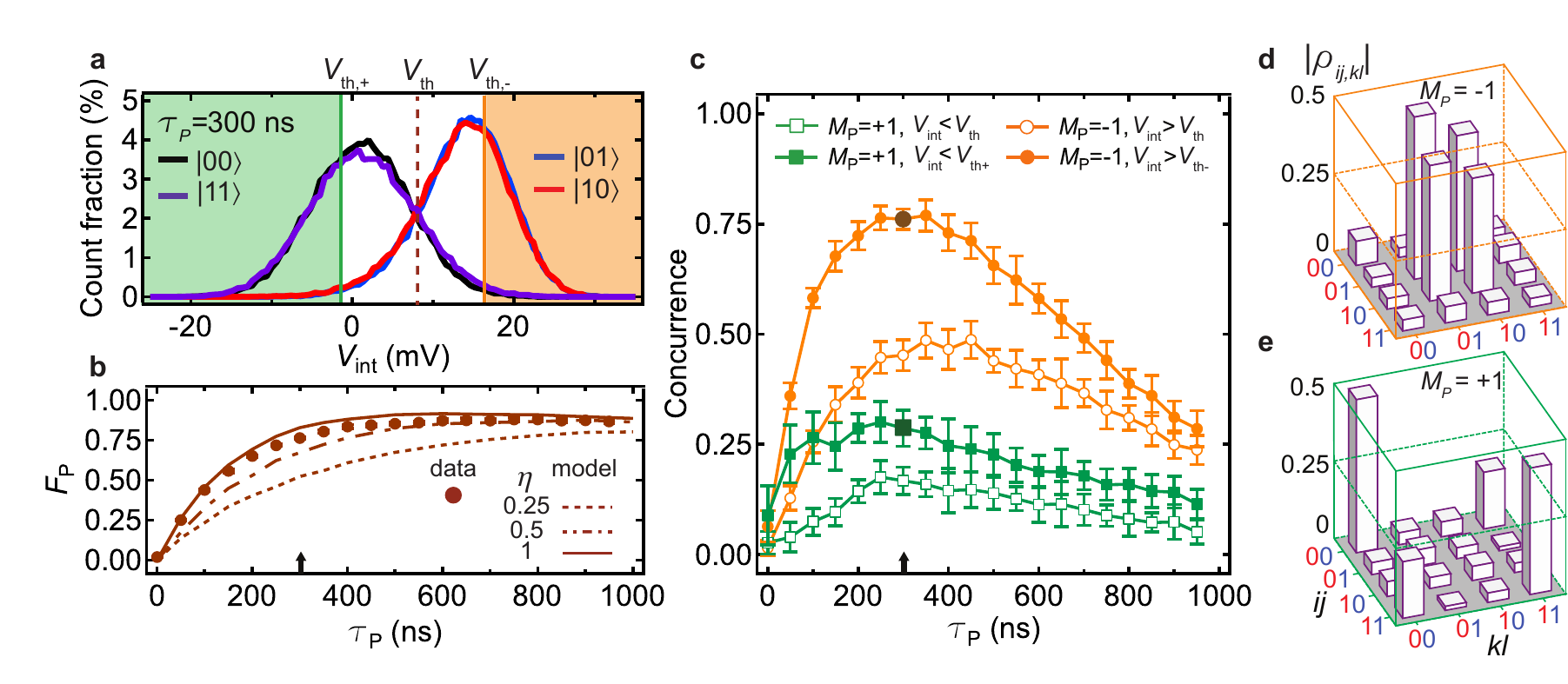}
\caption{{\bf Probabilistic entanglement generation by postselected parity measurement.} {\bf a}, Histograms of $\Vint$ ($\tp=300~\ns$) for the four computational states.  The results are digitized into $\Mp=1 (-1)$ for $\Vp$ below (above) a chosen threshold.  {\bf b,} Parity readout fidelity $\Fp$ as a function of $\tp$. We define $\Fp=1-\epse-\epso$, with $\epse=p(\Mp=-1|\mathrm{even})$ the readout error probability for a prepared even state, and similarly for $\epso$ (see also Fig.~S7). Data are corrected for residual qubit excitations (see Methods Summary). 
Error bars are smaller than the dot size. Model curves, shown for comparison, are obtained from $5,000$ quantum trajectories for each initial state and $\tp$, with quantum efficiencies $\eta=0.25,$ $0.5$, and $1$ for the readout amplification chain (see Methods Summary). {\bf c}, Concurrence $\conc$ of the two-qubit entangled state obtained by postselection on $\Mp=-1$ (orange) and on $\Mp=+1$ (green squares). Empty symbols correspond to the threshold $\Vthresh$ that maximizes $\Fp$, binning  $\psuccess \sim 50~\%$ of the data into each case. Solid symbols correspond to a threshold $\Vthreshm (\Vthreshp)$ for postselection on $\Mp=-1(+1)$, at which $\epso (\epse) = 0.01$. Concurrence is optimized at $\tp\sim300~\ns$, where  $\psuccess \sim 20\%$ in each case.
We employ maximum-likelihood estimation~\cite{Filipp09} (MLE) to ensure positive-semidefinite density matrices, but concurrence values obtained with and without MLE differ by less than $3\%$  over the full data set.
{\bf d,e,} State tomography conditioned on $\Vp>\Vthreshm$ (d) and $\Vp<\Vthreshp$ (e), with $\tp=300~\ns$, corresponding to the dark symbols in (c).}
\end{figure*}

\begin{figure*}
\includegraphics[width=1.9\columnwidth]{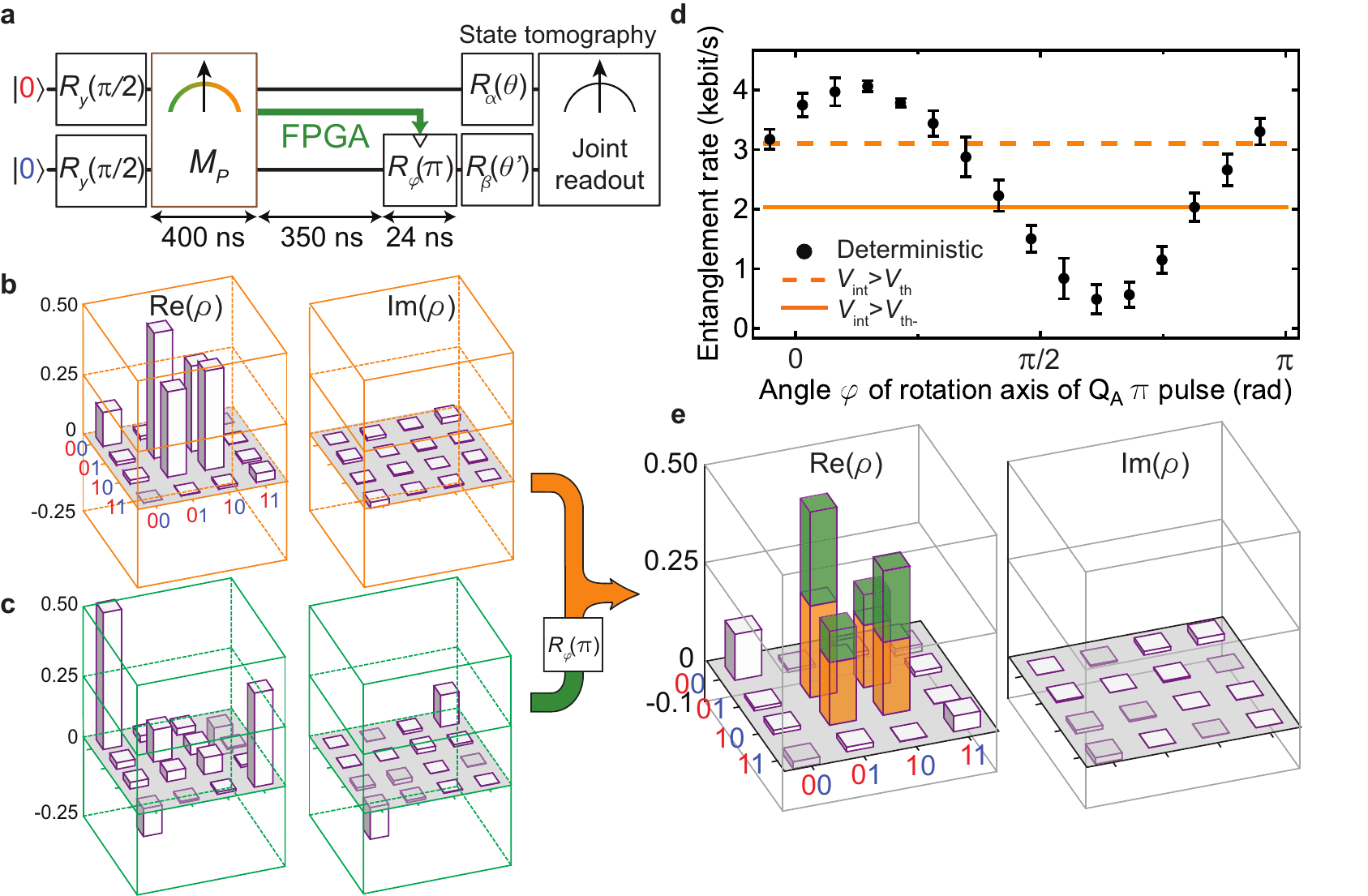}
\caption{{\bf Deterministic entanglement generation using feedback.} {\bf a},
We close a digital feedback loop by triggering (via the FPGA) a $\pi$ pulse on $\QA$ conditional on parity measurement result $\Mp=+1$. This $\pi$ pulse switches the two-qubit parity from even to odd, and allows the deterministic targeting of $\OddBellp=(\ket{01}+\ket{10})/\sqrt{2}$.
{\bf b,c,} Parity measurement results $\Mp=-1$ and $\Mp=+1$ each occur with $\sim50\%$ probability. Virtual $z$-gates compiled into the tomography pulses correct for the deterministic AC Stark phase acquired between $\ket{01}$ and $\ket{10}$ during parity measurement (due to residual mismatch between $\ChiA$ and $\ChiB$). A different AC Stark phase is acquired between $\ket{00}$ and $\ket{11}$, resulting in the state shown in (c), with the maximal overlap with even Bell state $[\ket{00}+\exp(-  i \varphi_\mathrm{e})\ket{11}]/\sqrt{2}$ at $\varphi_\mathrm{e}=0.73\pi$.
{\bf d,} Generation rate of entanglement using feedback, as a function of the azimuthal angle $\varphi$ of the $\pi$ pulse rotation axis. The deterministic entanglement generation rate outperforms the rates obtained with  postselection (Fig.~3). Error bars are the standard deviation of $7$ repetitions of the experiment at each $\varphi$. {\bf e}, Full state tomography for deterministic entanglement [$\varphi=(\pi-\varphi_\mathrm{e})/2$], achieving fidelity $\BraOddBellp \rho \OddBellp = 66\%$ to the targeted $\OddBellp$, and concurrence $\conc = 34\%$. Colored bars highlight the contribution from cases $\Mp=-1$ (orange) and $\Mp=+1$ (green).}
\end{figure*}

The ideal continuous parity measurement gradually suppresses the unconditioned density matrix elements $\rho_{ij,kl}=\bra{ij}\rho\ket{kl}$ connecting states with different parity (either $i\neq k$ or $j\neq l$), and leaves all other coherences (off-diagonal terms) and all populations (diagonal terms) unchanged.
The experimental tomography reveals the expected suppression of coherence between states of different parity (Fig.~2b-c). The temporal evolution of $|\rhoq_{11,10}|$, with near full suppression by $\tp=400~\ns$, is quantitatively matched by a master-equation simulation of the two-qubit system (see Methods Summary). Tomography also unveils a  non-ideality: albeit more gradually, our parity measurement partially suppresses the absolute coherence between equal-parity states, $|\rhoq_{01,10}|$ and $|\rhoq_{00,11}|$. The effect is also quantitatively captured by the model. Although intrinsic qubit decoherence contributes  (see Fig.~S4 for quantitative details), the dominant mechanism is the different AC-Stark phase shift induced by intra-cavity photons on basis states of the same parity~\cite{Lalumiere10,Tornberg10,Murch13}. This form of measurement back-action has both deterministic and stochastic components, and the latter suppresses absolute coherence under ensemble averaging.
We emphasize that this imperfection is technical rather than fundamental. It can be mitigated in the odd subspace by perfecting the matching of $\ChiB$ to $\ChiA$, and in the even subspace by increasing $\chi_\mathrm{A,B}/\kappa$ ($\sim 1.3$ in this experiment).

The ability to discern  parity subspaces while  preserving coherence within each opens the door to generating entanglement  by parity measurement on $\kpsii$. For every run of the sequence in Fig.~2, we discriminate $\Vint$ using the threshold $\Vthresh$ that maximizes the parity measurement fidelity $\Fp$ (Fig.~3a). Assigning $\Mp=+1\,(-1)$ to $\Vint$ below (above) $\Vthresh$, we bisect the tomographic measurements into two groups, and obtain the density matrix for each.  We quantify the entanglement achieved in each case using  concurrence $\conc$ as the metric~\cite{Horodecki09}, which ranges from $0\%$ for an unentangled state to $100\%$ for a Bell state. 
As $\tp$ grows (Fig.~3b), the optimal balance between increasing $\Fp$ at the cost of  measurement-induced dephasing and intrinsic decoherence is reached at $\sim300~\ns$ (Fig.~3c). Postselection on $\Mp=\pm1$ achieves $\conc_{|\Mp=-1} =45\pm 3\%$ and $\conc_{|\Mp=+1}=17\pm 3\%$, with each case occurring with  probability $\psuccess~\sim50\%$. The higher performance for $\Mp=-1$ results from lower measurement-induced dephasing in the odd subspace, consistent with Fig.~2.

The entanglement achieved by this probabilistic protocol can be increased with more stringent postselection. Setting a higher threshold
$\Vthreshm$ 
achieves $\conc_{|\Mp=-1} = 77\pm2\%$ but keeps $\psuccess\sim20\%$ of runs.
Analogously, using $\Vthreshp$ achieves $\conc_{|\Mp=+1}=29\pm4\%$ with similar $\psuccess$ (Figs.~3d,e). However, increasing $\conc$ at the expense of reduced $\psuccess$ is not evidently beneficial for QIP. For the many tasks calling for maximally-entangled qubit pairs (ebits), one may use an optimized distillation protocol~\cite{Horodecki09} to prepare one ebit from $N=1/\LN(\rho)$ pairs in a partially-entangled state $\rho$,
where $\LN$ is the logarithmic negativity~\cite{Horodecki09}. The net rate $\Ratee$ of ebit generation would be $\Ratee= \psuccess  \Rateexpt  \LN\left(\rhoq\right)$, 
where $\Rateexpt$ is the protocol repetition rate ($10~\kHz$ here). For postselection on $\Mp=-1$, we calculate $\Ratee=3.1~\kebits$ using $\Vthresh$ and $\Ratee=2.0~\kebits$ using $\Vthreshm$. Evidently, increasing entanglement at the expense of reducing $\psuccess$ is counterproductive in this context. 

Motivated by the above observation, we finally demonstrate the use of digital feedback control~\cite{Riste12b} to transform entanglement by parity measurement from probabilistic to deterministic, i.e., $\psuccess=100\%$. Using a 
FPGA controller, we apply a $\pi$ pulse on $\QA$ conditional on measuring $\Mp=+1$ (using $\Vthresh$, Fig.~4). In addition to switching the two-qubit parity, this pulse lets us choose which odd-parity Bell state to target by selecting the azimuthal angle of the pulse rotation axis. Clearly, we optimize deterministic entanglement by maximizing overlap to the same odd-parity Bell state for $\Mp=-1$ as for $\Mp=+1$. The highest deterministic $\conc=34\%$ achieved is lower than for our best probabilistic scheme, but the boost to $\psuccess=100\%$ achieves a higher $\rateE=4.1~\kebits$.

Our experiment extends the fundamental study of continuous measurement~\cite{Hatridge13, Murch13} in superconducting circuits to the multi-qubit scenario, providing a testbed for the investigation of wavefunction projection and induced dephasing.  
Furthermore, the implemented parity meter generates entanglement for any measurement result, making it suitable for deterministic QIP protocols. 
Specifically, the combination of parity measurement with digital feedback realizes the first multi-qubit measurement-based protocol in the solid state made deterministic through feedback, as achieved with photonic~\cite{Furusawa98},  ionic~\cite{Barrett04, Riebe04}, and atomic~\cite{Sherson06,Krauter13} systems. 
Future experiments will target the complementary use of analog feedback control to cancel the back-action caused by imperfections in the parity meter~\cite{Tornberg10, FriskKockum12}, making it robustly quantum nondemolition. This achievement will refine the mastery over quantum measurement and feedback~\cite{Devoret13} required to extend quantum coherence by active control methods.

\section{Methods Summary}
\subsection{Device parameters}
 Lorentzian best fits to cavity transmission (Fig.~1b) yield $\kappa=\kappa_{\mathrm{out}} +\kappa_{\mathrm{in}}=2 \pi \times (1.56\pm0.01~\MHz)$ and $\{\ChiA,\ChiB\}/\pi = \{-4.03\pm0.02,-4.21\pm0.02\}~\MHz$. 
From room-temperature characterization, we estimate asymmetric output/input couplings $\kappa_{\mathrm{out}}/\kappa_{\mathrm{in}}=8$. The qubits have transition frequencies $\{\fa,\fb\} = \{5.52,7.80\}~\GHz$, relaxation times $\{\ToneA,\ToneB\}= \{22,7\}~\us$, and pure dephasing times $\{\TtwoA,\TtwoB\} =\{11,8\}~\us$. 
Using the method detailed in Ref.~24\nocite{Riste12}, we estimate a residual excitation of $1\%(2\%)$ for $\QA(\QB)$.

\subsection{Readout signal processing}
In Fig.~1b we probe the cavity with a pulse ($\nss \sim 1.4$) at variable frequency, after preparing the qubits in one of the four computational states. The cavity transmission is acquired with homodyne detection at $10~\MHz$ intermediate frequency.
In Fig.~1c-d the cavity response ($\nss = 2.5$), first amplified by the JPA, is demodulated with $0$ intermediate frequency (measurement, local oscillator, and pump tones are provided by the same generator). 
For each shot, the average homodyne signal over a $2.5~\us$ window preceding state preparation is subtracted. This subtraction mitigates the infiltration of low-frequency fluctuations in the JPA bias.
In Figs.~2-4, $\ti=100~\ns$ and $\tf=\tp+150~\ns$, experimentally found to maximize $\Fp$. Similarly, an offset integrated over $2.5~\us$  is subtracted from each $\Vint$ (Fig.~S9).

\subsection{Model}
The system is described by the dispersive Hamiltonian~\cite{Blais04}
\begin{eqnarray*}
H/\hbar &=& \left(\omegar-\sum\limits_{q=\mathrm{A,B}} \chiQ\szq\right)a^{\dagger}a - \sum\limits_{q=\mathrm{A,B}}\frac{1}{2} \omegaq\szq \\
&&+\epsp \left[ a^{\dagger} e^{-i\omegap t} +  a  e^{+ i\omegap t}\right]\nonumber.
\end{eqnarray*}
The cavity-mediated qubit-qubit interaction  $J (\spB\smA + \smB \spA)$ is disregarded, as $J$ vanishes for $\ChiA=\ChiB$. 
We model the evolution of $\rhoq$ following the method of quantum trajectories in Refs.~8, 9, 21, 32. 
The stochastic master equation, valid for 
$t\ll \ToneA,\ToneB$,  is
\begin{eqnarray}
{\rm d}\rhoq&=& \frac{1}{i\hbar}\com{H}{\rhoq}\dt \nonumber\\
&&+\sum\limits_{q=\mathrm{A,B}} \left(\frac{1}{T_{1}^q} \mathcal{D}\left[\sigma_{-}^q \right] \rhoq+  \frac{1}{2\Ttwoq} \mathcal{D}\left[\szq\right]\rhoq\right) \dt \nonumber\\
&&- \sum\limits_{ijkl} \chi_{ij,kl} \left({\rm Im}\left[\alpha^*_{ij}\alpha_{kl}\right]+i{\rm Re}\left[\alpha^*_{ij}\alpha_{kl}\right]\right)\Pi_{ij}\rhoq\Pi_{kl}\dt \nonumber\\
&&+\sqrt{\kappa\eta} \mathcal{M}\left[\Pi_{\alpha}e^{-i\phi}\right]\rhoq{\rm d}W(t),\label{SDE}
\end{eqnarray}
with  operators  $\Pi_{ij}=\out{ij}{ij}$ and $\Pi_{\alpha}=\sum\limits_{ij} \alpha_{ij}\Pi_{ij}$, super-operators $\mathcal{D}\left[\Theta\right]\rhoq=\Theta\rhoq\Theta^{\dagger}-\frac{1}{2}\acom{\Theta^{\dagger}\Theta}{\rhoq}$ and  $\mathcal{M}\left[\Theta\right]\rhoq=\Theta\rhoq+\rhoq \Theta^{\dagger}-\expec{\Theta+\Theta^{\dagger}}\rhoq$. 
Here, $\phi$ is the homodyne-detection phase set by the JPA pump, and $\chi_{ij,kl}= \chi_{ij}-\chi_{kl}$, where $\chi_{ij} = \bra{ij}\sum\limits_{q=\mathrm{A,B}} \chiQ \szq \ket{ij}$. 
 The dynamics of $\alpha_{ij}$ in the frame rotating at  $\omegap$ is given by
\begin{eqnarray*}
\dot{\alpha}_{ij}=-i \epsp(t) - i\left(\omegar-\omegap+\chi_{ij}\right)\alpha_{ij}-\frac{\kappa}{2}\alpha_{ij}.~~~~
\end{eqnarray*}
$\mathrm{d}W$ is the noise in the homodyne record: 
\begin{eqnarray*}
\Vp(t) \dt \propto \sqrt{\kappa\eta} \langle \Pi_{\alpha} e^{-i\phi}+ \Pi_{\alpha}^\dagger e^{i\phi}\rangle {\rm d}t + {\rm d}W. \\\nonumber
\end{eqnarray*}
Quantum trajectories are unraveled by numerically solving Eq.~(\ref{SDE}) with $\dt=1~\ns$ and a Wiener white-noise process ${\rm d}W$ (zero mean, variance $\mathrm{d}t$) generated pseudo-randomly. For each trajectory, $\Vint$ is obtained using the same integration and offset-subtraction parameters as in the experiment. 
The unconditioned $\rhoq$ is obtained by solving Eq.~(\ref{SDE}) without the last term. ~~~~~~

\nocite{Gambetta08, Lalumiere10, Tornberg10, Hutchison09}

\section{Acknowledgments}
\begin{acknowledgments}
We thank  C.~C.~Bultink and H.-S.~Ku  for experimental assistance, as well as G.~Haack, R.~Hanson, G.~Johansson, A.~F.~Kockum, and L.~Tornberg for discussions. We acknowledge funding from the Dutch Organization for Fundamental Research on Matter (FOM), the Netherlands Organization for Scientific Research (NWO, VIDI scheme), the EU FP7 integrated projects SOLID and SCALEQIT, and partial support from the DARPA QuEST program.
\end{acknowledgments}

\section{Author contributions}
D.R. fabricated the device.  D.R. and C.A.W. performed the measurements. D.R., C.A.W. and G.d.L. analyzed the data. M.D., Ya.M.B. and L.D.C. provided theory support.  M.J.T. and R.N.S. realized the feedback controller. K.W.L. designed the JPA. D.R., G.d.L. and L.D.C. wrote the manuscript with feedback from all authors. L.D.C. designed and supervised the project.

{\bf Correspondence} and requests for materials should be addressed to L.D.C. (l.dicarlo@tudelft.nl).

\end{document}

\begin{figure*}
\includegraphics[width=0.5\columnwidth]{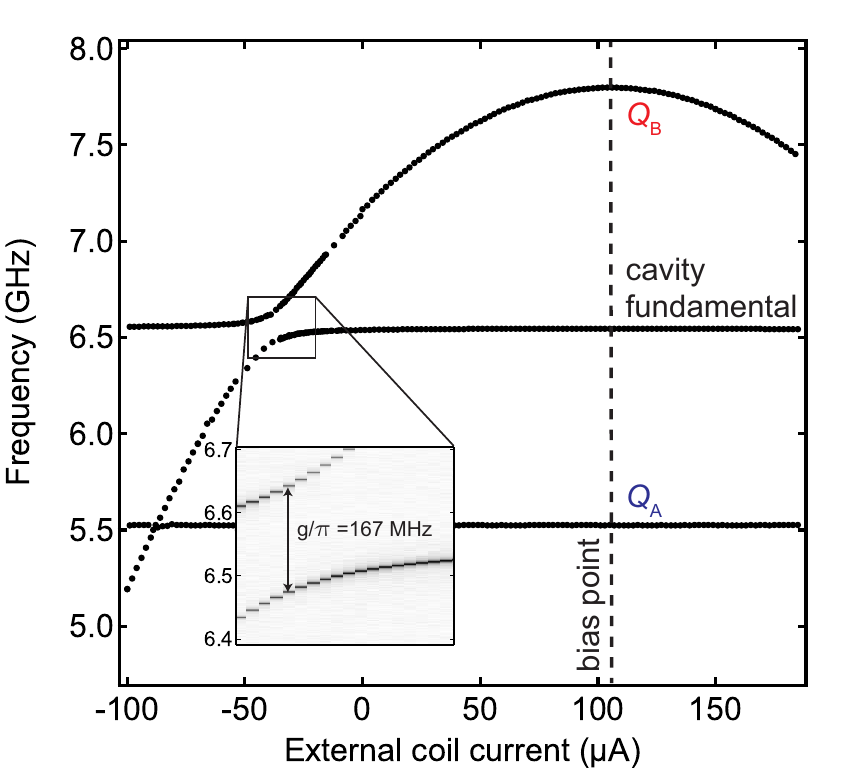}
\caption{{\bf Spectroscopy of the two-qubit and cavity system.} The transition frequency of $\QB$ is tuned by applying magnetic flux through its SQUID loop with an external coil. $\QA$ $(\fa = 5.52~\GHz)$ is a single-junction transmon and thus not tunable. $\QB$ was designed tunable to allow trimming of the dispersive-shift matching condition.  However, the maximal frequency of $\QB$ $(\fb = 7.80~\GHz$) is still approximately $20~\MHz$ lower than needed for a perfect match of dispersive shifts.  Thus, we flux bias $\QB$ at this maximal frequency, which is also optimal for coherence.
Inset: Higher resolution spectroscopy of the avoided crossing of $\QB$ with the cavity fundamental mode ($\fr=6.55~\GHz$), revealing a minimum splitting of 167 MHz.}
\end{figure*}

\begin{figure*}

\includegraphics[width=0.75\columnwidth]{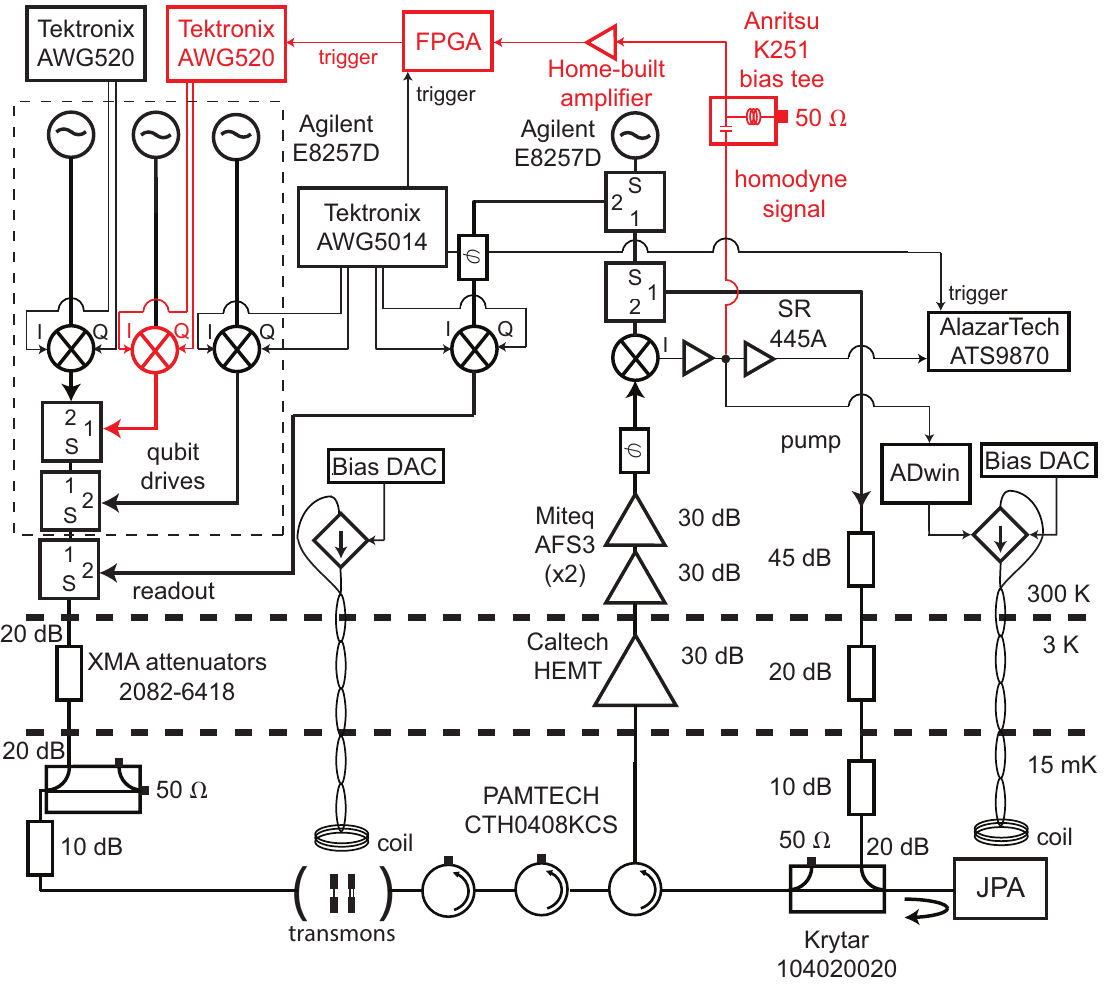}
\caption{{\bf Detailed schematic of the experimental setup.}
Complete wiring of electronic components outside and inside the $\Hethree/\Hefour$ dilution refrigerator (Leiden Cryogenics CF-650). 
Readout and qubit-drive pulses, shaped by a Tektronix AWG5014 and two AWG520, enter the cavity via a single transmission line.  The cavity output is reflected by the JPA, which is biased by a superconducting coil and a strong pump tone, bending its resonance down to $\fp$ and providing parametric amplification~\cite{Castellanos-Beltran08}. The signal is further amplified at the $3~\K$ stage (Caltech Cryo1-12, $0.06~\dB$ noise figure) and at room temperature (two Miteq AFS3-04000800-10-ULN amplifiers, $0.8~\dB$ noise figure). Demodulation to baseband is provided by a generator at $\fp$, also used for readout and pump. Two phase shifters allow adjusting the relative phase between the 3 tones at $\fp$.
The demodulated signal is split into three separate arms after amplification by a Stanford Research Systems SR445A.  One arm stabilizes the JPA flux bias via an ADwin-GOLD processor programmed as a PID controller~\cite{Riste12}. In the second arm, the signal is filtered by a bias tee, amplified with a home-built amplifier, and integrated and thresholded by the FPGA.  The FPGA conditionally triggers a $\QA$ $\pi$ pulse from an AWG520 (Fig. 4). The third arm connects to an AlazarTech ATS9870 digitizer for data storage and processing after a second SR445A amplification stage.}
\end{figure*}

\begin{figure*}

\includegraphics[width=0.5\columnwidth]{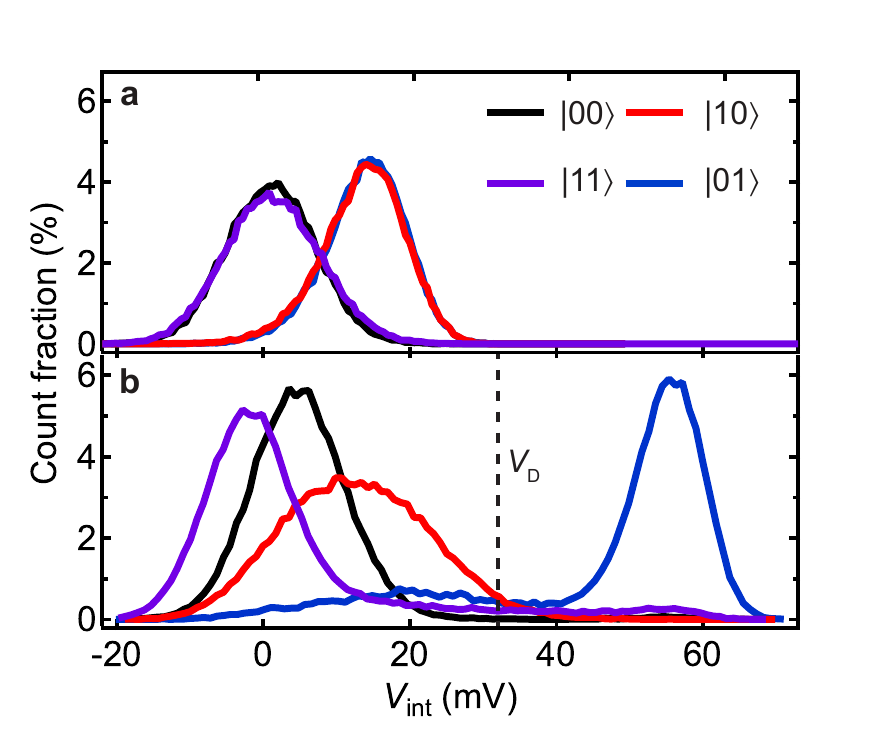}
\caption{{\bf Readout configuration for parity measurement and state tomography.}
{\bf a,} Histograms for the computational basis for the parity measurement $\Mp$ ($\tp~=~300~\ns, \nss=2.5$), as in Fig.~3a. At this measurement power, states within each parity subspace are largely indistinguishable (see also Fig.~1b and  Fig.~S7). For an ideal parity measurement, $\betaA = \betaB = 0$. We extract $\betaA = 0.0146~\mV, \betaB = -0.123~\mV, \betaBA = -6.25~\mV,$ and $\beta_0 = 7.46~\mV$. {\bf b,} Histograms for the tomography measurement (integration time $850~\ns$, $\nss\sim60$). At this power, the cavity response is nonlinear (critical photon number~\cite{Gambetta06} $n_\mathrm{crit}\sim60$), causing the resonance for $\ket{10}$ to bend towards lower frequency.  As the resonance for $\ket{01}$ is instead power-independent, this effect discriminates $\ket{01}$ from the other states.
This gives the joint readout the sensitivity to single and two-qubit terms required to perform state tomography~\cite{Filipp09}.  Averaging of raw tomography measurements yields  $\betaA = -8.10~\mV, \betaB = 9.10~\mV, \betaBA = -12.8~\mV,$ and $\beta_0 = 17.1~\mV$.  Digitizing the single shots with threshold $V_\mathrm{D} = 32~\mV$ gives  $\betaA = 0.424$, $\betaB = -0.360$, $\betaBA = 0.379$, and $\beta_0 = 0.540$.}
\end{figure*}

\begin{figure*}
\centering
\includegraphics[width=0.5\columnwidth]{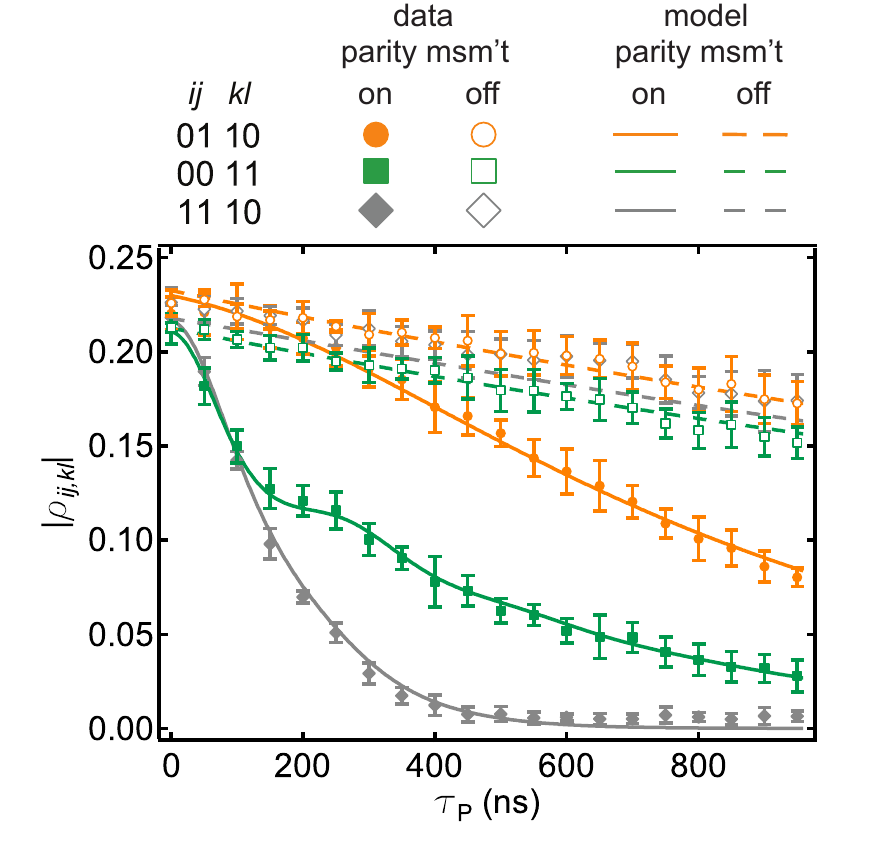}
\caption{{\bf Temporal evolution of two-qubit superposition state with and without continuous parity measurement.} Comparison of the unconditioned two-qubit evolution during  parity measurement (solid symbols, same data as in Fig.~2b) and during a delay of the same duration $\tp$ (empty symbols). In the latter case, the decay of $\absrhoijkl$ is solely due to intrinsic qubit decoherence.}
\end{figure*}

\begin{figure*}
\centering
\includegraphics[width=\columnwidth]{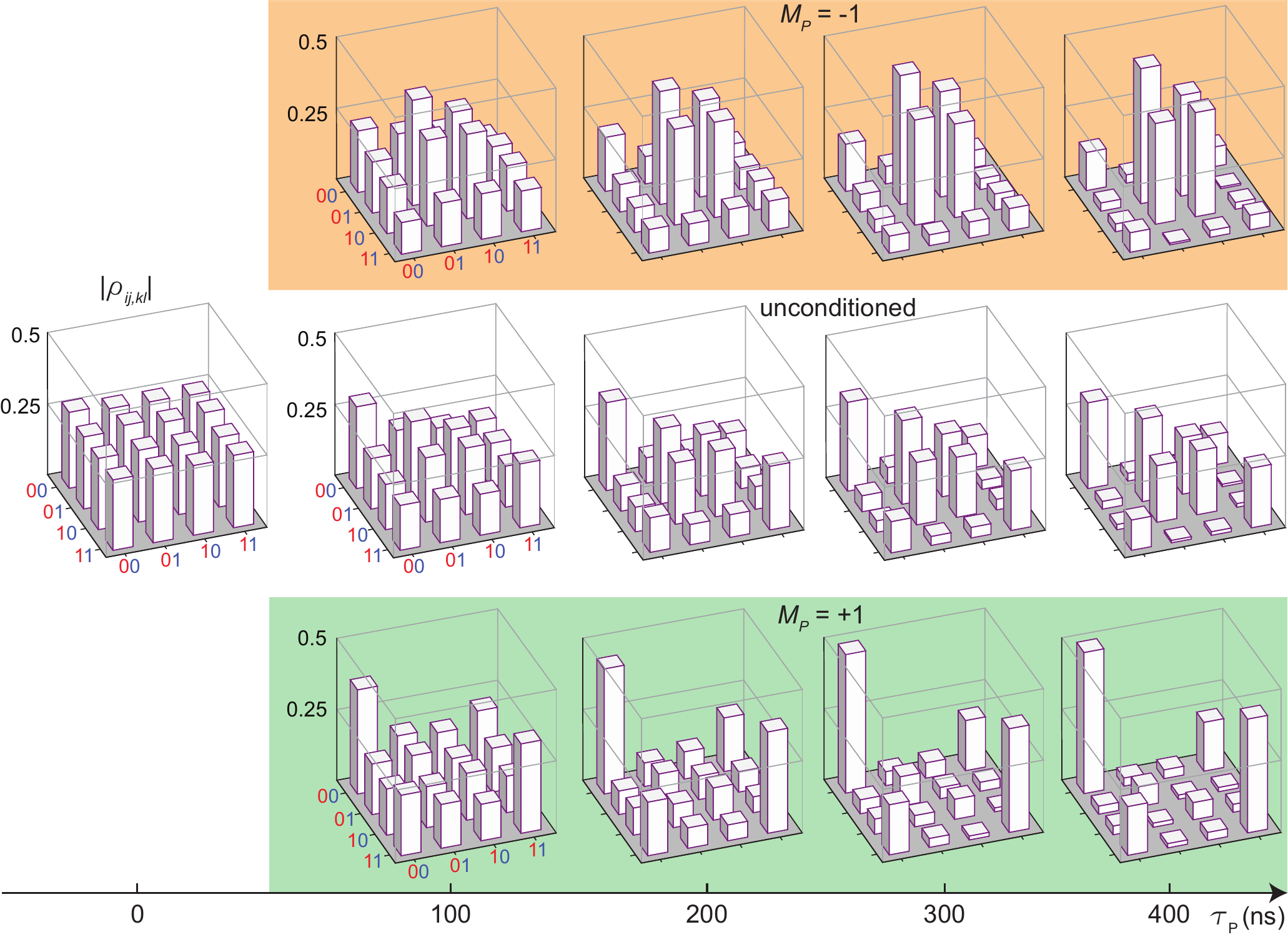}
\caption{{\bf Two-qubit evolution under continuous parity measurement.} Unconditioned and conditioned state tomography of the final two-qubit state similar to Figs.~2 and 3, but at more values of $\tp$ and using the threshold $\Vthresh$ optimizing parity readout fidelity ($\nss=2.5$). Middle row: unconditioned evolution. For $\tp=0$, there is only a $10~\ns$ buffer between state preparation and tomography, instead of the $350~\ns$ used in Figs.~2-4 and all other $\tp$ values here. The uniformity of $\absrhoijkl$ for $\tp=0$ ($<4\%$ relative difference) attests to the preparation fidelity of the initial maximal superposition state. Top row: evolution conditioned on $\Vint>\Vthresh$ ($\Mp=-1$), bottom row: evolution conditioned on $\Vint<\Vthresh$ ($\Mp=+1$).}
\end{figure*}

\begin{figure*}
\centering
\includegraphics[width=\columnwidth]{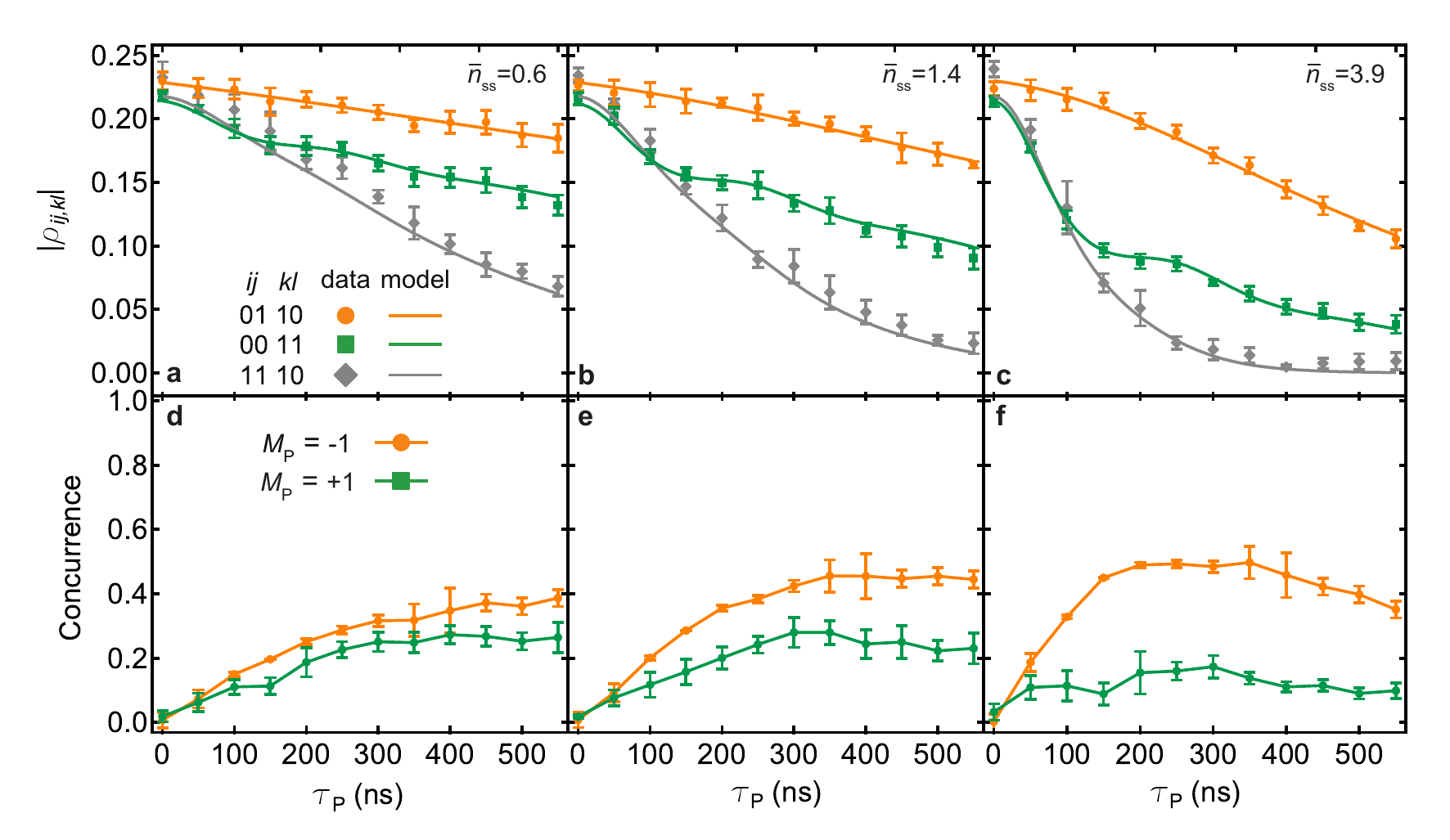}
\caption{{\bf Two-qubit unconditioned evolution and conditioned concurrence for different measurement strengths.} {\bf a,b,c,} Experiment as in Figs.~2-3 with measurement strength corresponding to $\nss=0.6\pm0.1$ (a,d), $1.4\pm 0.1$ (b,e), and $3.9\pm 0.1$ (c,f). 
The best-fit frequency mismatch $(\ChiA-\ChiB)/\pi$ (see also Fig.~S3) is $182\pm32~\kHz$ (a,d), $220\pm18~\kHz$ (b,e), and $275\pm7~\kHz$ (c,f). 
Concurrence is calculated after postselection on $\Vint<\Vthresh$ $(\Mp=+1)$ or $\Vint>\Vthresh$ $(\Mp=-1)$.}  
\end{figure*}

\begin{figure*}
\centering
\includegraphics[width=0.5\columnwidth]{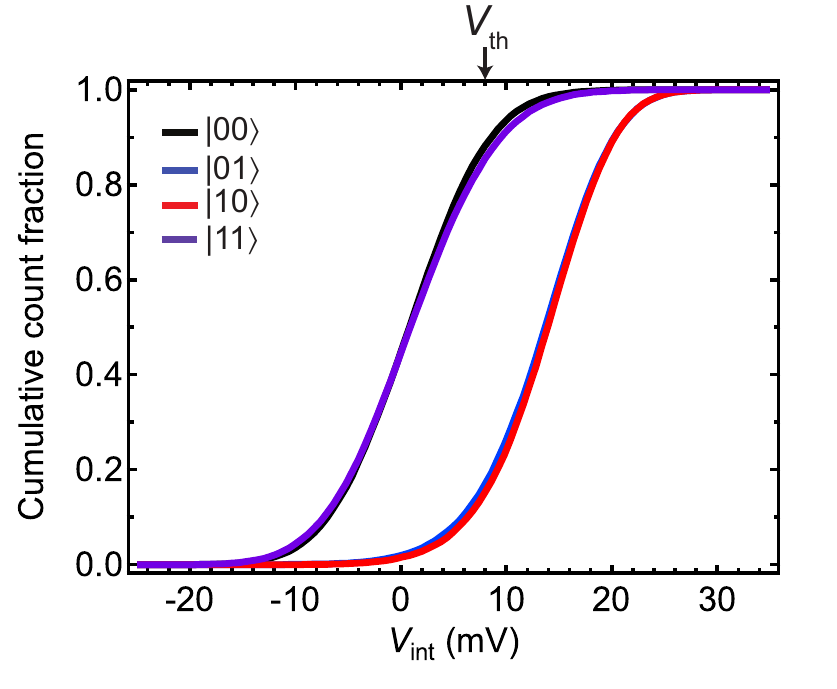}
\caption{{\bf Cumulative histograms of parity measurements.}
The four computational states are subjected to a parity measurement with $\tp = 300~\ns, \nss=2.5$, as in Fig.~3a. At the optimal threshold $\Vthresh$ (dashed line), the average errors in determining the parity are $\epse = 0.13, \epso = 0.11$, yielding a parity measurement fidelity of $\Fp=1-\epse-\epso=0.76$ (corrected for residual qubit excitations, see Methods Summary).  In a similar manner, we define the distinguishability within each parity subspace as the fidelity of the measurement discriminating between those states, yielding $0.03$ for the even subspace and $0.02$ for the odd.  }  
\end{figure*}

\begin{figure*}
\centering
\includegraphics[width=0.5\columnwidth]{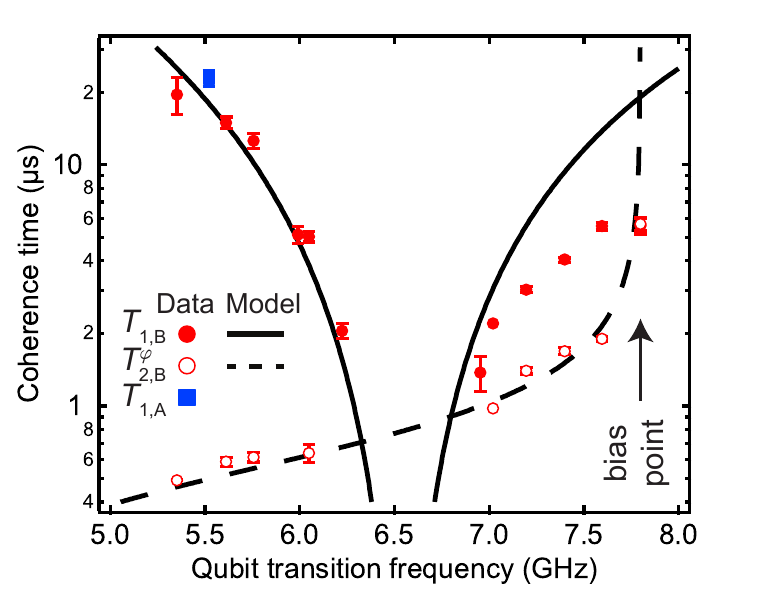}
\caption{{\bf Frequency-dependent coherence times of $\QB$.}
Energy relaxation times $\ToneB$ (filled circles) and $\ToneA$ (square) below the fundamental cavity resonance are consistent with the single-mode Purcell effect~\cite{Houck08} and a coupling strength $g/\pi = 167~\MHz$ at the $\QB$-cavity avoided crossing, as extracted from spectroscopy  (Fig.~S1). We attribute the lower $\ToneB$ above the fundamental resonance to the effect of higher cavity modes.  
Pure dephasing times $\TtwoB$ (open circles) are in excellent agreement with the first-order approximation for flux noise~\cite{Koch07} with spectral density $S_f\left(\omega\right) = A^2/|f|$ and best-fit $A = 1.9\pm0.1 \cdot 10^{-5} \Phi_0$  (dashed line), with $\Phi_0$ the flux quantum.}
\end{figure*}

\begin{figure*}
\centering
\includegraphics[width=0.5\columnwidth]{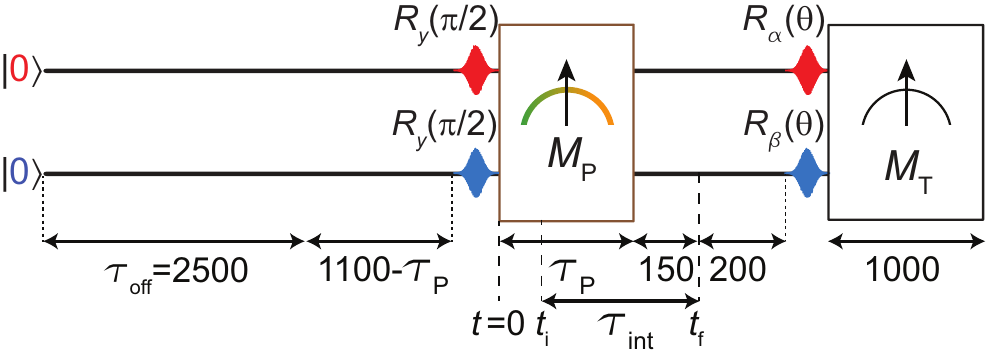}
\caption{{\bf  Pulse timing and measurement integration windows.}
Extended view (not to scale)  of the pulse sequence used in Figs.~2-4, showing also the integration windows used for parity measurement ($\tint$ for signal and $\toff$ for offset) and tomographic joint readout. All specified time intervals are expressed in $\ns$. Qubit control is performed with DRAG pulses~\cite{Motzoi09} with Gaussian envelopes on the main quadrature ($\sigma=6~\ns$, $4\sigma$ total duration) and derivative-of-Gaussian envelopes of optimized amplitude on the other. Single-qubit pulses are applied sequentially ($\QB$ first), with $10~\ns$ buffer between them. The tomography measurement pulse is $1~\us$ long, and the homodyne response integrated for $850~\ns$ starting after the first $100~\ns$. }
\end{figure*}